\documentclass[aps,epsfig,prl,preprint,nofootinbib]{revtex4}
\usepackage[dvips]{graphicx}
\usepackage[dvips]{color}
\usepackage{subfigure}
\usepackage{epsfig}
\usepackage{dcolumn} 
\usepackage{bm} 
\usepackage{amsmath}
\usepackage{amssymb}
\newcommand\beq{\begin{equation}}
\newcommand\bear{\begin{eqnarray}}
\newcommand\eeq{\end{equation}}
\newcommand\eear{\end{eqnarray}}

\begin{document}

\title{
Rigidity in Condensed Matter and Its Origin in Configurational Constraint
}
\author{Shibu Saw and Peter Harrowell}
\affiliation{School of Chemistry, University of Sydney, Sydney NSW 2006 Australia}

\date{\today}

\begin{abstract}{
Motivated by the formal argument that a non-zero shear modulus is the result of averaging
over a constrained configurations space, we demonstrate that the shear modulus calculated
over a range of temperatures and averaging times can be expressed (relative to its infinite
frequency value) as a single function of the mean squared displacement. This result is shown
to hold for both a glass-liquid and a crystal-liquid system.
}
\end{abstract}

\pacs{xx}

\maketitle

\noindent  Glasses are rigid and liquids are not. The difficulty with any distinction of two phases based
solely on rigidity is that the property is not an equilibrium one. Over 45 years ago, Lebowitz\cite{lebowitz}
and Ruelle\cite{ruelle} pointed out that, in the thermodynamic limit, the free energy of a phase
cannot depend on the shape of the sample and so the equilibrium value of the shear modulus
must vanish for all phases $-$ crystals as well as glasses $-$ in the limit of large N. Rationalising
the obvious point that rigid materials do in fact exist, a number or researchers\cite{frenkel-sausset} have
concluded that a non-zero shear modulus is a property of a metastable state and hence rigidity
is observable only for observation times shorter than the lifetime of that state. Since the
observation of a non-zero shear modulus depends crucially on this lifetime, it would seem
that any theoretical treatment of the mechanical properties of a material will depend on
solving the onerous problem of slow relaxation in a condensed phase. Williams and Evans\cite{evans},
acknowledging this difficulty, suggested that the shear modulus be formally calculated as
an equilibrium average over a constrained space of configurations. This perspective suggests
the attractive possibility that the magnitude of the shear modulus might be expressed as an
explicit function of the magnitude of the configurational constraint applied, a relation that
includes a threshold degree of constraint, below which rigidity vanishes. In this paper we
establish just such a relationship between the shear modulus and the configurational
constraint, measured here by the mean squared displacement, for both a glass-liquid and
crystal-liquid system.

\vspace{5mm} \noindent    The Squire-Holt-Hoover expression\cite{squire} for the (constrained) equilibrium shear modulus $G_{eq}$
of a solid is
\begin{eqnarray}
G_{eq} = G_{\infty} -\beta V [<\sigma^2>-<\sigma>^2]
\end{eqnarray}
where $\sigma$ is the shear stress, $\beta=1/k_BT$, $V$ is volume and $G_{\infty}$ is the infinite frequency (or Born)
shear modulus given by\cite{Ilg}
\begin{eqnarray}
G_{\infty} = \frac{N}{V} k_BT - \frac{1}{2V} \sum_i \sum_{j \ne i}  < \Big( y_{ij}^2 F_{ij}\Big[1 - \frac{x_{ij}^2}{r_{ij}^2} \Big ]  -\frac{d^2\phi}{dr_{ij}^2} \frac{x_{ij}^2 y_{ij}^2}{r_{ij}^2}  \Big ) >
\end{eqnarray}
where $F_{ij} = -\dfrac{1}{r_{ij}}\dfrac{d\phi}{dr_{ij}}$ and $\phi(r)$ is a spherically symmetric inter-particle potential.
Note that the shear modulus $G_{eq}$ is reduced, relative to the high frequency value, by an amount
associated with variance of the shear stress fluctuations. In the context of elastic theory, these
fluctuations correspond to non-affine contributions to the modulus\cite{Lemaitre-Maloney}. What is measured in a
typical experiment is the stress relaxation function $G(t)=\sigma(t)/\gamma$, where $\gamma$ is a applied strain
and $\sigma(t)$  is the resulting time dependent shear stress. The relation between $G(t)$ and $G_{eq}$ is
given by the following expression\cite{yoshino1},
\begin{eqnarray}
G(t) = G_{eq} + \beta V [<\sigma(0)\sigma(t)> - <\sigma>^2 ]
\end{eqnarray}
where the shear stress autocorrelation function $ < \sigma(0)\sigma(t) > $ equals $ < \sigma^2> $ when t=0 and $<\sigma>^2$ in the limit $t\rightarrow \infty$.
It follows from Eq. 3 that $G_{eq}$ represents a lower bound to the
observed modulus $G(t)$ with $\lim\limits_{t \to \infty} G(t)=G_{eq}$. This long time limit refers only to the {\it explicit} time
dependence arising from the shear stress autocorrelation function. It does not include any
{\it implicit} time dependence associated with the observation time used to construct the averages
in $G_{eq}$ (see Eq. 1). So, the averages $<...>$ in Eq. 1-3 are understood to be taken over some
constrained configuration space. In the absence of a constraint, $<\sigma>=0$ and $G_{\infty}=\beta V <\sigma^2>$\cite{zwanzig-mountain}
so that $G_{eq}=0$.

\vspace{5mm}\noindent   The model liquid used in this study is a 2D system of soft disks with a pair interaction
potential, $\phi_{ij}(r)=\epsilon \Big( \dfrac{a_{ij}}{r} \Big)^{12} $, between species $i$ and $j$.
In the case of the binary equimolar
mixture we use $a_{11} =1.0$, $a_{22} =1.4$ and $a_{12} =1.2$ and all particle with unit mass, a model that has
been extensively studied\cite{perera} in the context of the glass transition. The temperature is reported in units of $\epsilon/k_{B}$  and time in units of $\tau=\sqrt{ma_{11}^2/\epsilon}$.
Simulations were
carried out under constant NVT conditions using LAMMPS\cite{lammps} with a Nose-Hoover
thermostats at reduced densities $0.7468$ (binary mixture) and $1.398$ (single component) with a
potential cut-off distance of $6.3 a_{11}$ . The system consisted of a total of N = 1024 particles in
the case of the binary mixture and $N = 1400$ for the single component system. Previously\cite{abraham},
we established that these values of N were sufficient for accurate calculation of the
stress fluctuations. At low temperatures, the trajectories are non-ergodic for all accessible
values of the averaging time $t$. In order to fairly sample the configuration space at these low temperatures we have averaged trajectories over statistically distinct initial configurations.  For the
binary mixtures, $51$ uncorrelated configurations were generated by cooling a liquid,
equilibrated at $T = 0.60$, to $T = 0.30$ at a cooling rate of $5 \times 10^{-5}$ and then minimizing the
potential energy of the resulting $T=0.30$ liquid by conjugate gradient minimization. The
minima, referred to as inherent structures (IS), were statistically independent as established
by the average shear stress (at $T = 0$) of the inherent structures equalling zero. To calculate
$G_{eq}$ and $G_{\infty}$ at a given temperature $T$ we randomly assigned momenta from the Boltzmann
distribution consistent with a temperature $T$ to the particles in each of the n IS configurations
and then determined the moduli for each individual IS using Eqs. 1and 2, respectively, by
averaging over a trajectory run for a time interval t. To obtain our final values of $G_{eq}$ and $G_{\infty}$,
we averaged the moduli for the individual IS configurations over all $51$ IS configurations. In
the case of the single component system, this protocol was modified as follows. For the
crystal phase we only used a single inherent structure, that of the perfect crystal. For the
liquid phase data we simply carried out averages over MD trajectories of the equilibrated
liquid.

\begin{figure}[t]
\vspace{4mm}
\includegraphics[scale=0.45,angle=0]{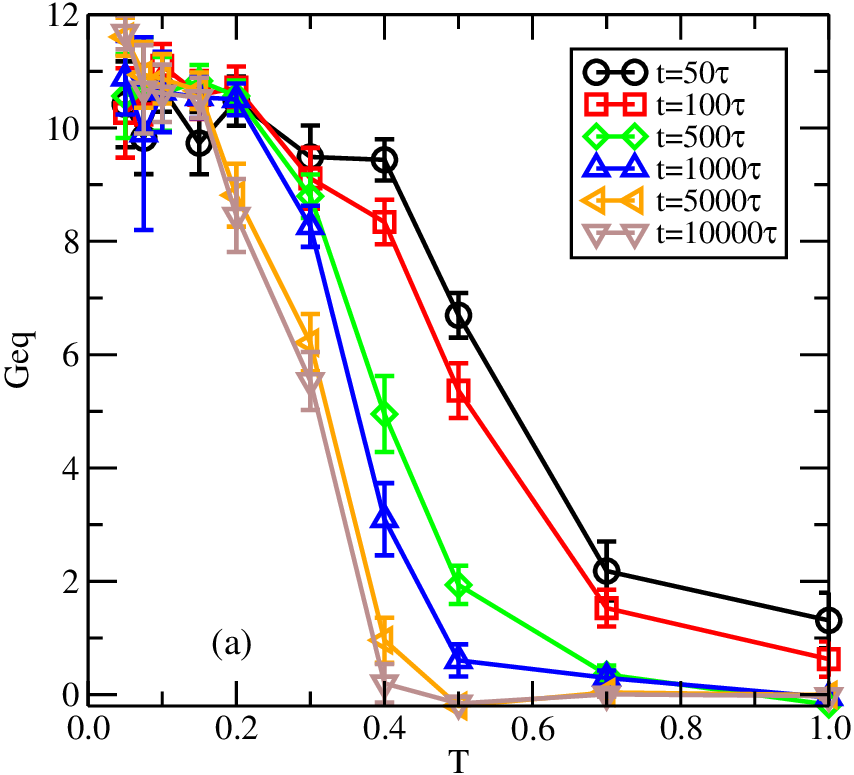}\hspace{18mm}
\includegraphics[scale=0.45,angle=0]{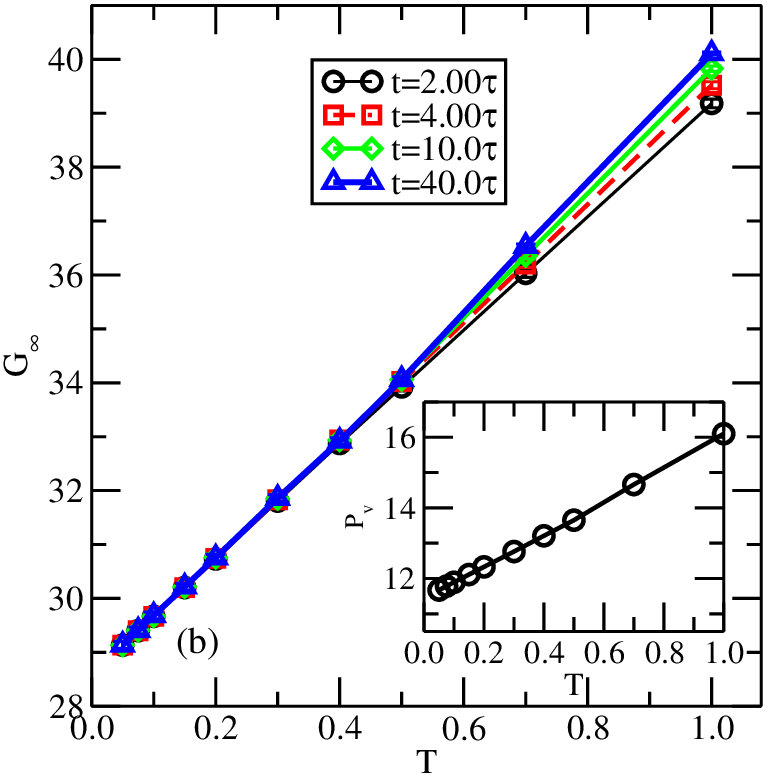}
\caption{(Color online)
Plot of (a) $G_{eq}$ and (b) $G_{\infty}$ vs $T$ for the binary mixture over different averaging
times. Note the significant effect of the averaging time in the case of $G_{eq}$ in contrast to $G_{\infty}$,
were the influence of averaging time has saturated within a short time $\sim 40\tau$. Insert in Fig.~1b: The virial
pressure $P_{\mathrm{v}} = -\dfrac{1}{2V} < \sum\limits_i \sum\limits_{j \ne i} r_{ij} \dfrac{d\phi}{dr_{ij}} > $ as a function of $T$.
}
\label{Fig1}
\end{figure}

\vspace{5mm}\noindent  In Fig.~1 we plot the values of $G_{eq}$ and $G_{\infty}$ as a function of $T$ for a binary mixture of soft disks
in $2D$ at a fixed density. We find that $G_{\infty}$ increases linearly with $T$ and shows no significant
variation with the averaging time t. (Note that the infinite frequency modulus referred to
experimentally is not $G_{\infty}$ but the value of $G(t)$ in the plateau region.) The equilibrium
modulus $G_{eq}$, in contrast, exhibits a strongly nonlinear decrease with increasing temperature,
to finally vanish at a sufficiently high temperature. The family of curves presented in Fig.~1 is
evidence of the significance of the time t used to calculate the statistics of the stress
fluctuations.

\vspace{5mm}\noindent   A number of papers\cite{barrat,wittmer,zaccone1} have discussed the loss of rigidity of a glass as characterised by
the disappearance of $G_{eq}$ on heating in the context of a possible thermodynamic instability,
analogous to the softening in a superheated crystal\cite{barrat,wittmer}, or as an un-jamming transition
associated with the thermal expansion of the amorphous solid\cite{zaccone1}. (We remind the reader
that our calculations have been carried out at fixed density so that this latter proposal is not
directly relevant here.) While the role of the observation time is discussed\cite{wittmer}, it is the
temperature that is treated as the essential control parameter for the transition. Following on
from our opening discussion, we shall explore the idea that this decrease of the equilibrium
shear modulus, either through increasing T or the observation time, is most fundamentally
expressed as a result of the changing degree of configurational constraint associated with the
averaging.

\vspace{5mm}\noindent    To begin we note that temperature dependence of $G_{\infty}$ is not associated with configurational
constraint since, by construction, the infinite frequency modulus depends only on the
sampling of the local curvature of the potential energy surface. The increase in $G_{\infty}$ with
temperature at constant volume demonstrated in Fig.~1b is directly associated with the
increase of the virial pressure with T under the constant volume constraint (see insert Fig.~ 1b). To
eliminate this additional temperature dependence we shall therefore consider the reduced
modulus $G_{eq}/G_{\infty}$. Next, we need a measure of the configurational constraint. The simplest
such measure is the particle mean squared displacement,
\begin{eqnarray}
<\Delta r^2(t)> =\frac{1}{N} \sum_{i} < |\overset{\rightarrow} r_{i}(0) - \overset{\rightarrow} r_{i}(t) |^2 >
\end{eqnarray}
where the time $t$ here is the same as the observation time used to calculate the stress averages
and, in the case of a binary mixture, the average is over both species.

\begin{figure}[t]
\vspace{4mm}
\includegraphics[scale=0.45,angle=0]{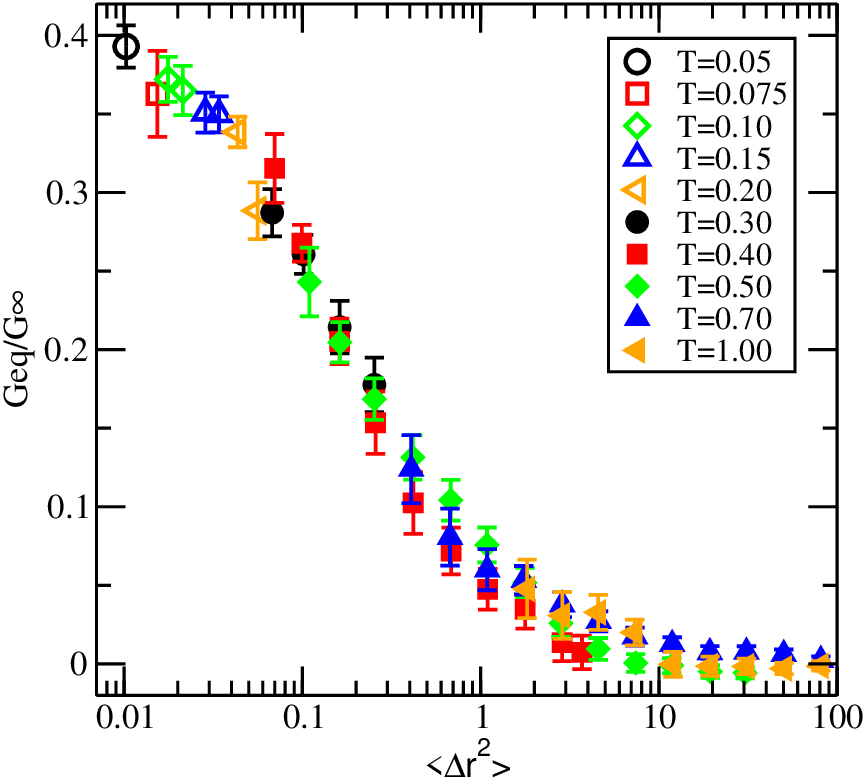}
\caption{(Color online)
Plot of $G_{eq}/G_{\infty}$ vs $<\Delta r^2>$ for the binary mixture. In each case, the mean squared
displacement is calculated over the same time interval as that use to evaluate $G_{eq}$.
}
\label{Fig2}
\end{figure}

\vspace{5mm}\noindent   In Fig. 2 we plot $G_{eq}/G_{\infty}$ vs $<\Delta r^2>$ where we have used the data from Fig.~1 for a range of
temperatures and observation times. We find that all of the data from Fig.~1 collapses onto a
single curve. This result provides strong support the twin propositions of this paper, i.e. that
the (reduced) shear modulus is simply a consequence of configurational constraint and that
the mean squared displacement provides a useful measure of this constraint.

\begin{figure}[b]
\vspace{4mm}
\includegraphics[scale=0.45,angle=0]{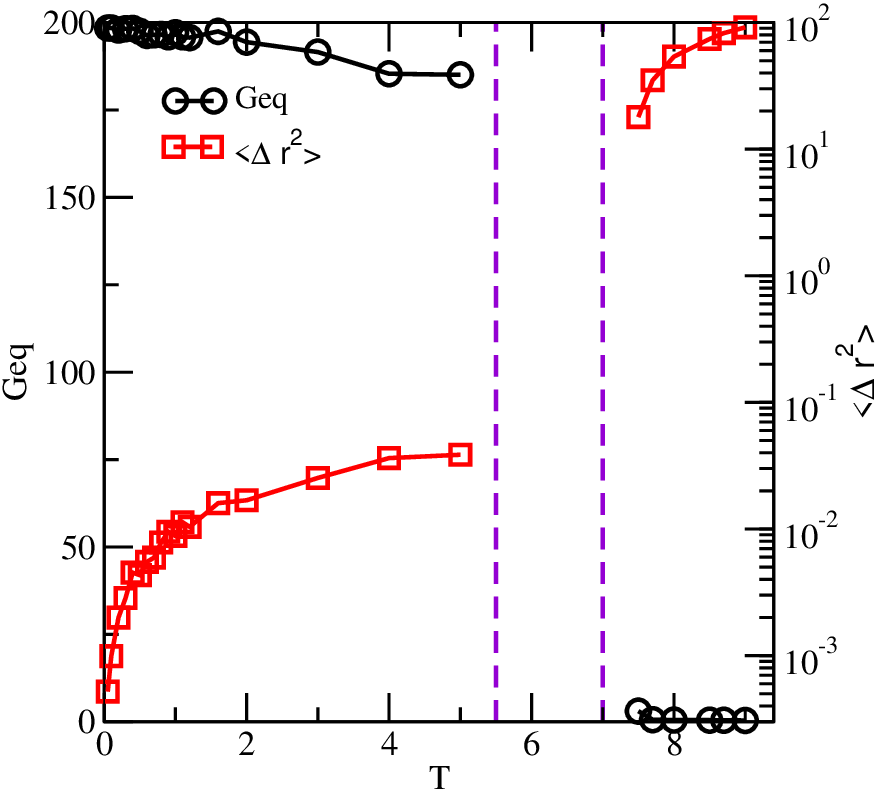}
\caption{(Color online)
The temperature dependence of $G_{eq}$ and $<\Delta r^2>$  (both averaged over a time $199\tau$) for
the single component 2D soft disk system. Due to the constant density constraint, there is a
range of temperatures corresponding to a two phase coexistence, indicated by the two vertical
dashed lines. The freezing transition is marked by step-like change in the modulus and the
mean squared displacement.
}
\label{Fig3}
\end{figure}

\vspace{5mm}\noindent   A glass forming liquid is convenient for our purposes because it can access the entire range of
$G_{eq}$ without encountering a thermodynamic singularity. Our argument relating shear modulus
and configurational constraint, however, should apply equally to crystallizing liquids. To
demonstrate this point, we consider a single component soft disk liquid in 2D which
crystallizes readily into a triangular lattice. In Fig.~3 we plot the values of $G_{eq}$ and  $<\Delta r^2>$
for the system as a function of $T$, using an observation time $t=199\tau$. The presence of the $1^{st}$
order freezing at $T =5.0$ is clearly evident in both quantities. In Fig.~4 we plot  $G_{eq}/G_{\infty}$ vs $<\Delta r^2>$
for a range of temperatures (using crystal and liquid configurations for $T$ below and
above $T_m$, respectively) and a range of observation times. Again, we find the data collapsed
onto a common curve, this in spite of the discontinuity of the modulus and $<\Delta r^2>$ with
respect to temperature. It is worth noting the striking difference in the low $T$ limit of  $G_{eq}/G_{\infty}$
for the crystal (Fig.~4) and the glass (Fig.~2). The reason for the considerable softening of the
glass relative to $G_{\infty}$ even at $T = 0$ is due to (i) the higher density of crystal and (ii) the large
non-affine motions\cite{Lemaitre-Maloney} in the glass relative to those in the crystal. The presence of non-affine
motions in the amorphous phase and their effective absence in the crystal is due to the
absence of inversion symmetry in the local structure of the amorphous phase and its presence
in the crystal\cite{zaccone2}.

\vspace{5mm}\noindent   The dependence of the reduced shear modulus $G_{eq}/G_{\infty}$ on $<\Delta r^2>$ is found, empirically, to
be well described by the following relation,
\begin{eqnarray}
\frac{G_{eq}}{G_{\infty}} = \frac{G_{eq}}{G_{\infty}}\Big|_{T=0.05}  \exp\Big(-q \ln^{\alpha} \Big[ \frac{<\Delta r^2>}{<\Delta r^2>\Big|_{T=0.05} }\Big]   \Big)
\end{eqnarray}

\vspace{5mm}\noindent    The success of this function is shown in Fig.~5 for the glass forming mixture with the fitted
values $\alpha=2$ and $q=0.08$. Eq.~5 also provides an excellent fit to $G_{eq}/G_{\infty}$ vs $<\Delta r^2>$ for both
the crystal and liquid phases of the single component system (see Fig.~4), but with different
parameters, $\alpha=2.9$ and $q = 0.0061$. Viewed as an emergent property of restricted particle
fluctuations, the derivation of the dependence of  $G_{eq}/G_{\infty}$ on the degree of configurational
constraint must represent a problem of fundamental importance.

\begin{figure}[t]
\vspace{4mm}
\includegraphics[scale=0.45,angle=0]{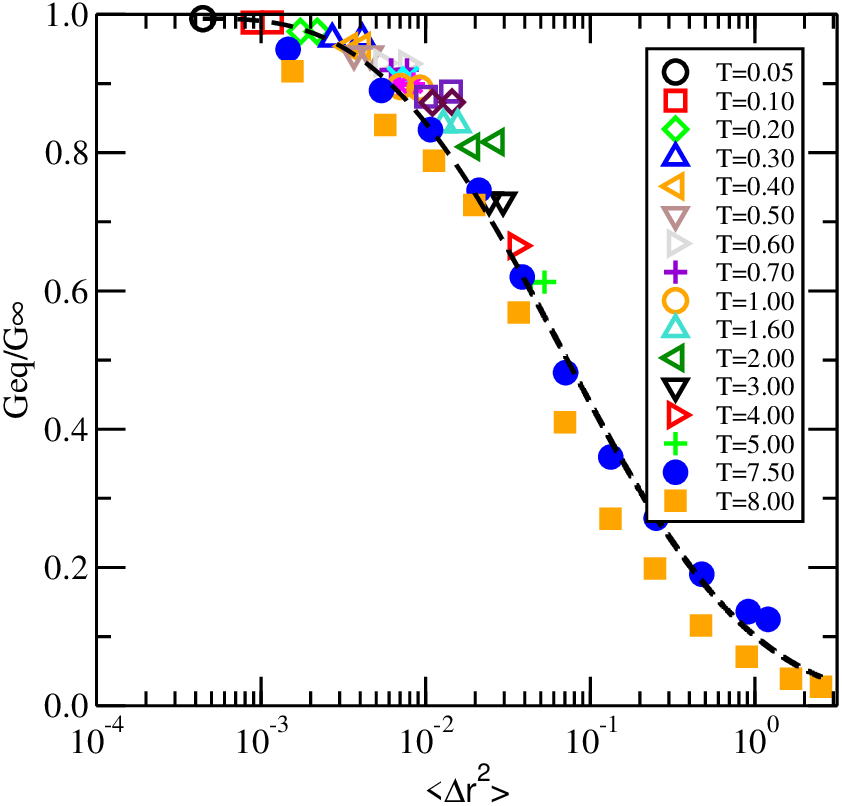}
\caption{(Color online)
The dependence of $G_{eq}/G_{\infty}$ on $<\Delta r^2>$  for the single component crystal and liquid.
As for Fig.~2, each point corresponds to a choice of $T$ and the averaging time. Liquid state
data i.e. $T > 5.0$  are presented by filled symbols and crystal data by open symbols. The curve
corresponds of a fit of Eq.~5 to the single component data with $\alpha=2.9$ and $q = 0.0061$.
}
\label{Fig4}
\end{figure}

\vspace{5mm}\noindent  We have argued here that the value of $G_{eq}$ (relative to $G_{\infty}$) is a consequence of constraint.
This is the opposite to the account provided within harmonic models of solids in which the
elastic constants (or the bond force constants) are prescribed in the model and the mean
squared displacement are determined as a consequence. This latter treatment, however, is
only possible because of the implicit configurational constraints (i.e. assumed elasticity,
unbreakable harmonic bonds, etc.) on which such models rely. For the harmonic solid,
$G_{eq} <\Delta r^2>/T =\mathrm{constant}$ (at fixed density). As shown in Fig.~5 (insert), this relation holds
only for $<\Delta r^2> <0.1$, a result that underscores the inclusion of anharmonic effects in the
empirical relations demonstrated in Fig.~4 and 5. Yoshino and Zamponi\cite{yoshino2} have recently
derived a power law relationship between the shear modulus and the mean squared
displacement in a granular model that applies within a metabasin, a restricted range of
configuration space corresponding, roughly, to  $0.01 \le <\Delta r^2> \ge 0.2$.
\begin{figure}[b]
\vspace{4mm}
\includegraphics[scale=0.45,angle=0]{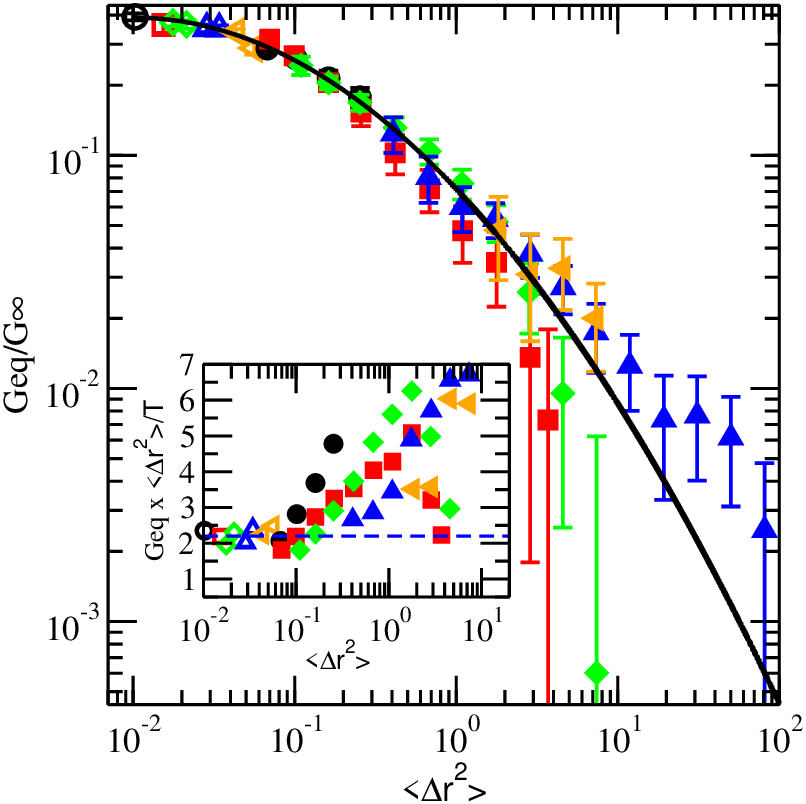}
\caption{(Color online)
A log-log plot of of $G_{eq}/G_{\infty}$ vs $<\Delta r^2>$ for the binary mixture. (The correspondence
between symbols and temperatures are the same as in Fig.~2). The expression in Eq.~5 (solid
curve) provides a good description over the entire range of $<\Delta r^2>$ with $\alpha=2.0$ and $q=0.08$.
Insert: The quantity $G_{eq} <\Delta r^2>/T $ vs $<\Delta r^2>$. The harmonic approximation,
indicated by a constant value (dashed line), breaks down for $<\Delta r^2> \ge 0.1$.
}
\label{Fig5}
\end{figure}

\vspace{5mm}\noindent In conclusion, we have verified that our two propositions: (1) the degree of configurational
constraint determines the magnitude of the shear modulus (relative to the low temperature
limit), and (2) the mean squared displacement provides a useful measure of this constraint,
do indeed represent a consistent physical picture for both a glass forming liquid and one that
undergoes freezing. This result represents a fundamental unification of the physical basis of
rigidity. The presence of a non-zero shear modulus is not, we argue, the consequence of a low
temperature, a high frequency measurement or even the presence of long range order. Rather,
each of these factors is important only in as far as they contribute to an implicit constraint on
the volume of configuration space that can be explored by stress fluctuations. It is this
constraint, however it is achieved, that determines the value of the equilibrium shear
modulus. This is a powerful result with a number of interesting consequences. First, accounts
of the temperature dependence of the shear modulus of metallic glasses\cite{mitrofanov} have relied on
the language of anharmonic effects borrowed from crystal physics. In the picture we present
here, the decrease in the glass modulus on heating is associated the increase in $<\Delta r^2>$ by
harmonic or anharmonic motions (along with any decrease in $G_{\infty}$ associated with thermal
expansion when a constant pressure is employed as in ref.~\cite{wittmer} ). Second, we have argued that
the shear modulus should be regarded as a mechanical manifestation of restricted motion.
Couple this idea with a description of the role that elastic behaviour plays in determining the
rate of particle motion (e.g. the shoving model of Dyre\cite{dyre}) and there is possibility of a self
consistent theory in which the modulus is, itself, a consequence of the very particle mobilities
that it acts to constrain. Thirdly, these results suggest a reassessment of the empirical
Lindemann criterion\cite{lindemann}, i.e. the observation that crystal order is lost once the mean squared
displacement exceeds some threshold value. Our results here suggest that it is rigidity, not
structure {\it per se}, that vanishes as the mean square displacement increases. Finally, since our
account of rigidity places no special condition on how the configuration space is accessed it
is possible that non-thermal contributions to particle mobility such as the non-affine motion
due to applied strain should result in an analogous reduction in $G_{eq}/G_{\infty}$\cite{saw}. Each of these
lines of inquiry is currently under investigation.

\vspace{5mm}\noindent In this paper we have established that the collection of factors $-$ time, temperature and order $-$
associated with the observation of rigidity in a dense phase can be replaced by a single
tangible length that characterizes the degree of configurational constraint. While we have
established that $<\Delta r^2>$ provides a workable measure of this constraint length, further work is
required to establish whether there is a better measure of this constraint and whether we can
derive from first principles the mathematical relationship between this measure and the shear
modulus.

\section{Acknowledgements}
We acknowledge support from the Australian Research Council.


\end{document}